\documentclass[12pt]{article}
\usepackage{jheppub}
\usepackage{amsmath}
\usepackage{amsfonts} 
\usepackage{color}
\usepackage{float}
\usepackage{slashed} 
\usepackage{graphicx}
\usepackage{graphics} 
\newcommand{\be}{\begin{equation}}
\newcommand{\ee}{\end{equation}}
\newcommand{\bea}{\begin{eqnarray}}
\newcommand{\eea}{\end{eqnarray}}
\newcommand{\ba}[1]{\begin{array}{#1}} 
\newcommand{\ea}{\end{array}}
\newcommand{\sign}{\text{sgn}} 
\title{\boldmath Fermionic correlators on the holographic neutron star}
\author{Lucas Acito,}
\author{Tob\'ias Canavesi,}
\author{Nicol\'as Grandi,}
\author{and Adri\'{a}n Lugo }
\affiliation{{\it Instituto de F\'\i sica La Plata (IFLP), CONICET \& 
Departmento de F\'\i sica, UNLP\\
C.C. 67, (1900) La Plata, Argentina.}}
\emailAdd{lucasacito@gmail.com}
\emailAdd{canavesi@fisica.unlp.edu.ar}
\emailAdd{grandi@fisica.unlp.edu.ar}
\emailAdd{lugo@fisica.unlp.edu.ar}
   
\abstract{
We investigate the fermionic perturbations of the asymptotically AdS background known as \emph{holographic neutron star}, which represents a highly degenerate state of strongly coupled fermions on a sphere at finite temperature. We calculate the two-point correlator of a fermionic operator and obtain its scaling properties as we approach the critical region of the phase diagram. }

\begin{document}
\maketitle

%\newpage
%
\section{Introduction}
Understanding the phenomenon of  superconductivity with high critical temperature is a prominent open problem in modern condensed matter theory. The phase diagram of the superconducting compounds exhibits a rich set of common features, including the presence of a superconducting dome at intermediate dopings and low temperatures, limiting at higher temperatures with an unconventional metallic phase \cite{Sachdev:2009}. The properties of this \emph{strange metal} are attributed to strong quantum fluctuations originating from a quantum critical point at zero temperature beneath the dome \cite{Hartnoll:2009sz}.

High temperature superconductivity's strong coupling nature makes it a well suited arena to test and employ holographic methods. In this context, the standard holographic description of the strange metallic phase, known as the \emph{electron star}, is a gravitational background consisting of a planar asymptotically AdS spacetime sourced by a charged perfect fluid \cite{Hartnoll:2010gu,Hartnoll:2010xj,Hartnoll:2011dm}. 
The fluid occupies an infinitely deep sea, that fills the bulk up to a finite value of the holographic coordinate. Closer to the boundary, spacetime is empty and the geometry approaches that of AdS in the Poincar\'e patch. The boundary theory represents a fermionic condensate at finite chemical potential. Electron stars can also be studied at finite temperature \cite{Hartnoll:2010ik,Puletti:2010de}. In such case, the fluid sea has a finite depth in the holographic direction, and there is a horizon beyond it. 

However, the underlying conformal invariance of the boundary theory makes it difficult to isolate the temperature $T$ from the chemical potential $\mu$ (doping) axes in the corresponding phase diagram. Indeed, the relevant observables are functions of the single dimensionless quotient $\mu/T$.
This dependence can be disentangled by the introduction of an additional scale $s$ that allows for the dimensionless ratios $T/s$ and $\mu/s$ to enter separately in the calculated quantities. 
 In \cite{Kiritsis:2015hoa}, such scale was introduced via a second chemical potential, given as the boundary value of an additional bulk $U(1)$ field. This technique allows for a two-dimensional phase diagram that can then be compared with that of high temperature superconducting materials. As expected, for certain parameter choices, there is a superconducting dome at low temperatures and intermediate doping that limits to a high-temperature strange metallic phase \cite{Kiritsis:2015hoa,Giordano:2018bsf}.
  
A different way to introduce a scale is to place the boundary system on a finite volume vessel. In this context, the \emph{holographic neutron star} originally introduced in \cite{deBoer:2009wk,Arsiwalla:2010bt} results in a boundary theory contained on the surface of a two-dimensional sphere. From the bulk perspective, it is sourced by a spherical droplet of a neutral perfect fluid, and it asymptotes global AdS. Its finite temperature extensions can also be studied \cite{Arguelles:2017pmx}, and the resulting $\mu$ versus $T$ phase diagram can be plotted. It presents an unstable region at intermediate doping \cite{Arguelles:2019mxh}, where the system shows a set of critical features. In particular, power-law dependencies appear in the density profile at the boundary of the star, as well as in the frequency dependence of the scalar two-point correlator \cite{Canavesi:2021aoh}. See the PhD thesis \cite{Canavesi:2023sqx} for a comprehensive review. 

In this paper, we explore further the holographic neutron star at finite temperature. We focus on the fermionic two-point correlator and its behavior as we approach the critical region. In Section \ref{sec:neutronstar}, we provide a brief introduction to the holographic neutron star geometry. Since it is well-discussed in the literature \cite{deBoer:2009wk,Canavesi:2021aoh}, we postpone any technical detail to Appendix \ref{app:neutronstar}. In Section \ref{sec:fermions}, we sketch the calculation of the fermionic two-point correlator. This is a straightforward adaptation of the corresponding calculations for the electron star case \cite{Hartnoll:2011dm}, which is explained in Appendices \ref{app:DiracopS2}, \ref{app:DiracopAdS}, \ref{app:correlator} and \ref{app:WKB}. Our results are presented and discussed in Section \ref{sec:results}.

\newpage
\section{The holographic neutron star}
\label{sec:neutronstar}
\begin{figure}\label{fig:phase.diagram}
	\begin{center}
	\includegraphics[width=12cm]{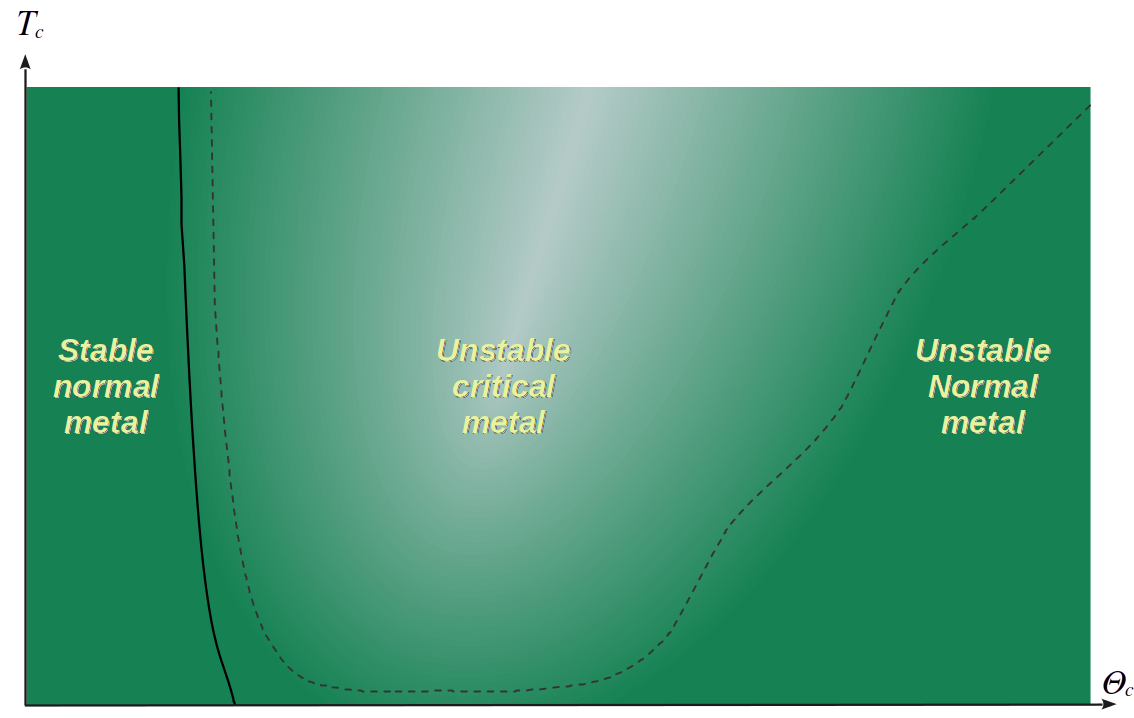}
		\caption{Phase diagram of the holographic neutron star at finite temperature. The unstable critical metal is characterized by power law behaviour of a variety of physical observables. 
Remarkably, such region is wider for larger temperatures which is reminiscent of the known phenomenology for high temperature superconductors.}
	\end{center}
\end{figure}

The holographic neutron star is defined as a neutral perfect fluid droplet that back-reacts into global AdS spacetime. For the explicit construction see Appendix \ref{app:neutronstar} and the original references \cite{deBoer:2009wk, Arsiwalla:2010bt}. Its ground state is given by a spherically symmetric  metric
\be\label{eq:metric.main}
ds^2 =  L^2\left(-e^{\nu(r)} dt^2+ e^{\lambda(r)}dr^2 + r^2d\Omega^2\right)\,,
\ee
which is sourced by the energy momentum tensor of the perfect fluid. The energy density and pressure correspond to those of a non-interacting gas in the Thomas-Fermi approximation. Einstein equations for this system turn into a set of Tolman-Oppenheimer-Volkoff equations, which are then solved numerically for the functions $\nu$ and $\lambda$. The star has a boundary, and at large distance from it the geometry asymptotes that of global AdS. The dual field theory corresponds to a highly degenerate state consisting of strongly interacting fermions confined to a two dimensional spherical vessel. The solutions are indexed by the central chemical potential $\mu_c$, which in the boundary theory plays the role of a doping axis. 
A temperature axis $T_c$ can be straightforwardly included by performing the Thomas-Fermi approximation in the bulk at finite temperature. This results in a very interesting phenomenology \cite{Arguelles:2017pmx}, summarized on the phase diagram of Fig.\,\ref{fig:phase.diagram}. 

Performing a Katz stability analysis in the grand canonical ensemble, and a turning point analysis in the microcanonical ensemble, the phase diagram can be divided into three regions \cite{Arguelles:2019mxh}. In the {\em stable normal metal} region, the system is fully stable according to both criteria. In the {\em unstable normal metal} region, Katz analysis detects one unstable eigenmode. Finally, in the {\em unstable critical metal} region, Katz criterion shows two unstable eigenmodes, and the turning point criterion signals an instability. 

Regarding the density profiles, as $\mu_c$ (or equivalently the central degeneracy $\Theta_c =(\mu_c-{\sf m})/T_c$ where $\sf m$ is the fermion mass) is increased at fixed $T_c$, they develop a  dense  core and a diluted halo, similar to what was previously known for the flat space astrophysical case \cite{Ruffini:2014zfa}. Interestingly, in the unstable critical metal region the edge of the halo shows a power law behaviour as a function of the radius. The same happens for the edge of the core, in the cases in which it is present. 

The two-point correlator for scalar perturbations enriches the diagram as follows. In the stable normal metal phase, the frequency dependence of the correlator is dominated by its poles. As we move to the right into the unstable normal metal region and then into the unstable critical metal region, the poles get separated and the correlator gets dominated by a power law contribution \cite{Canavesi:2021aoh}. Moreover, in the large mass limit, the two-point scalar correlator as a function of the angular separation at equal times develops a swallow tail structure. % in the unstable critical region.

\newpage
\section{Fermionic correlators}
\label{sec:fermions}
To study fermionic perturbations on the neutron star background, we introduce a Dirac spinor with mass ${\sf m}$ representing the fermionic fluctuations, and solve the corresponding Dirac equation. The angular part has eigenfunctions with half-integer indices $j$ and $m$ with $|m|\leq j$ and a sign index $\epsilon=\pm 1$, see Appendix \ref{app:DiracopS2} for details. They have a purely harmonic time dependence with frequency $\omega$. To obtain the radial dependence, we must solve an effective Schr\"odinger equation  
\be
\label{eq:Schroedinger}
\phi''_{\omega  jm\epsilon }(r) - U_{\omega  j \epsilon }(r)\;\phi_{\omega jm\epsilon }(r)=0\,.
\ee
The precise form of the Dirac spinor in terms of the wavefunction $\phi_{\omega  jm\epsilon }(r)$, as well as the details of the potential $U_{\omega  j \epsilon }(r)$ are discussed in the Appendix \ref{app:DiracopAdS}. %The relevant point here is that, depending on the values of $j$ and $\omega$, the potential may have either two turning points or none. 

This spinor bulk field is dual to a fermionic boundary operator ${\cal O}_{\omega jm\epsilon}$, whose two-point correlator ${\cal G}_{\omega jm\epsilon}=\langle {\cal O}_{\omega jm\epsilon}{\cal O}_{\omega jm\epsilon}\rangle$ we want to calculate. To do that, we have to solve for the 
Schroedinger wavefunction with regular boundary conditions at the interior, and then approach the AdS boundary to obtain the expansion
\be
\label{eq:culoculoasintotico`}
\phi_{\omega jm\epsilon} (r)
\approx
A_{\omega jm\epsilon} \left(r^{1+\epsilon {\sf m}L}+\dots\right)+
B_{\omega jm\epsilon} \left(r^{-\epsilon{\sf m}L}+\dots\right)\,.
\ee
The fermionic two-point correlators  can then be written in terms of the leading $A_{\omega jm\epsilon}$ and subleading $B_{\omega jm\epsilon}$ coefficients, in the form
%
%\be\label{eq:correlators}
%{\cal G}_{\omega jm+} = \left( \frac{{2\sf m}L +1}{\omega+\mu+\left|j+\frac12\right|}\,\right)^{\!2} \!{\cal G}_{\omega jm-} = L^3\,\frac{{2\sf m}L +1}{\omega+\mu+\left|j+\frac12\right|}\,
%\frac{B_{\omega jm+}}{A_{\omega jm+}}\,. 
%\ee 
%
\be\label{eq:correlators}
{\cal G}_{\omega jm-} = \left( \frac{\omega+\mu+\left|j+\frac12\right|}{{2\sf m}L +1}\,\right)^{\!2} \!{\cal G}_{\omega jm+} = L^3\,\frac{\omega+\mu+\left|j+\frac12\right|}{{2\sf m}L +1}\,
\frac{B_{\omega jm+}}{A_{\omega jm+}}\,. 
\ee 
The details of the calculation of the correlator are explained in Appendix \ref{app:correlator}.

We take the limit of large fermion mass ${\sf m}L\gg1$, keeping the combinations $E=\omega /{\sf m}L$ and $J=(j+1/2)/{\sf m}L$ constant. They become real continuous labels, in terms of which the potential can be written as $U_{\omega j\epsilon}(r)\underset{{\sf m}L\rightarrow \infty}{\rightarrow}{\sf m}^2L^2\;V_{E J}(r)$, with
\be\label{eq:WKBpot}
V_{E J}(r) =e^{\lambda(r)} \left(1+\frac{J^2}{r^2}-\frac{E^2}{e^{\nu(r)}}\right) 
\,.
\ee
In this limit the wavefunction in \eqref{eq:Schroedinger}
 can be solved analytically using the WKB approximation, the detailed calculations are provided in  Appendix \ref{app:WKB}. For the values of $E$ and $J$ for which the potential has two turning points, the fermionic correlators read
%
%\be
%\label{eq:correlators.WKB}
%{\cal G}_{EJ+} 
%=
%-
% \frac{L^3 r_2^{2{\sf m}L}}{E+J} \,
%e^{-2{\sf m}L\int_{r_2}^{\infty} \frac {dr}r\left(
%\sqrt{r^2V_{EJ}(r)}-1\right)}
%\,\tan\!
%\left(\!
%{\sf m}L\!\!\int_{r_1}^{r_2}\!\! dr' \sqrt{-V_{EJ}(r')}
%\right)\,.
%\ee
\be
\label{eq:correlators.WKB}
{\cal G}_{EJ-} 
=
-
 \frac{L^3 }{4}r_\Lambda^{-2{\sf m}L}(E+J) \,
e^{-2{\sf m}L\int_{r_2}^{r_\Lambda}  {dr}  
\sqrt{ V_{EJ}(r)}}
\,\tan\!
\left(
{\sf m}L\!\int_{r_1}^{r_2}\! dr' \sqrt{-V_{EJ}(r')}
\right)\,,
\ee
where $r_\Lambda$ is an UV cutoff.
From this expression we can easily obtain the position of the poles, which determine the normal mode energies $E_n(J)$  according to the Bohr-Sommerfeld quantization condition
\be 
\label{eq:bohr.sommerfeld}
\left.{\sf m}L\int_{r_1}^{r_2} dr'\,\sqrt{-V_{E J}(r')}\;\right|_{E=E_n(J)} 
= \left(n + \frac{1}{2}\right)\,\pi\,,
\quad \quad n\in\mathbb{N}_0\,.
\ee
Moreover, we can obtain an expression for the residua $r_n(J)$ around each pole, whose explicit form reads
\begin{equation}
r_n(J)= \frac{L^2 }{4{\sf m}}\,\frac{E_n(J)+J}{E_n(J)} \,
\frac{e^{-2{\sf m}L\int_{r_2}^{r_\Lambda}  {dr}  
\sqrt{ V_{EJ}(r)}}}{
\int_{r_1}^{r_2} dr'\,\frac{ e^{\lambda(r)-\nu(r)}}{\sqrt{-V_{E_n(J)J}}}}\,r_\Lambda^{-2{\sf m}L}\,.
\label{eq:residua0}
\end{equation}
More details are given in Appendix \ref{app:WKB}. This allows us to define a \emph{regular part} of the correlator, by substracting the poles as in 
\begin{equation}
\label{eq:regular}
{\cal G}_{EJ-}^{\sf reg}={\cal G}_{EJ-}-\sum_n\frac{r_n(J)}{E-E_n(J)}\,.
\end{equation}

\newpage

Having the expressions for the pole positions \eqref{eq:bohr.sommerfeld} and the regular part of the correlator \eqref{eq:regular}, we want to answer the following questions:
\begin{enumerate}
\item Do the poles get separated as we approach the critical region?
\item Does the regular part take a power law form there?
\end{enumerate}
If the answer to both questions is affirmative, we can safely state that the two-point correlator becomes dominated by its power law component in the critical region.

We investigate these issues in what follows, by numerically evaluating equation \eqref{eq:bohr.sommerfeld} in {\tt Mathematica}, and complementarily by solving \eqref{eq:Schroedinger} using a shooting method, in the different regions of the phase diagram.

\begin{figure}[t]
\includegraphics[width=\textwidth]{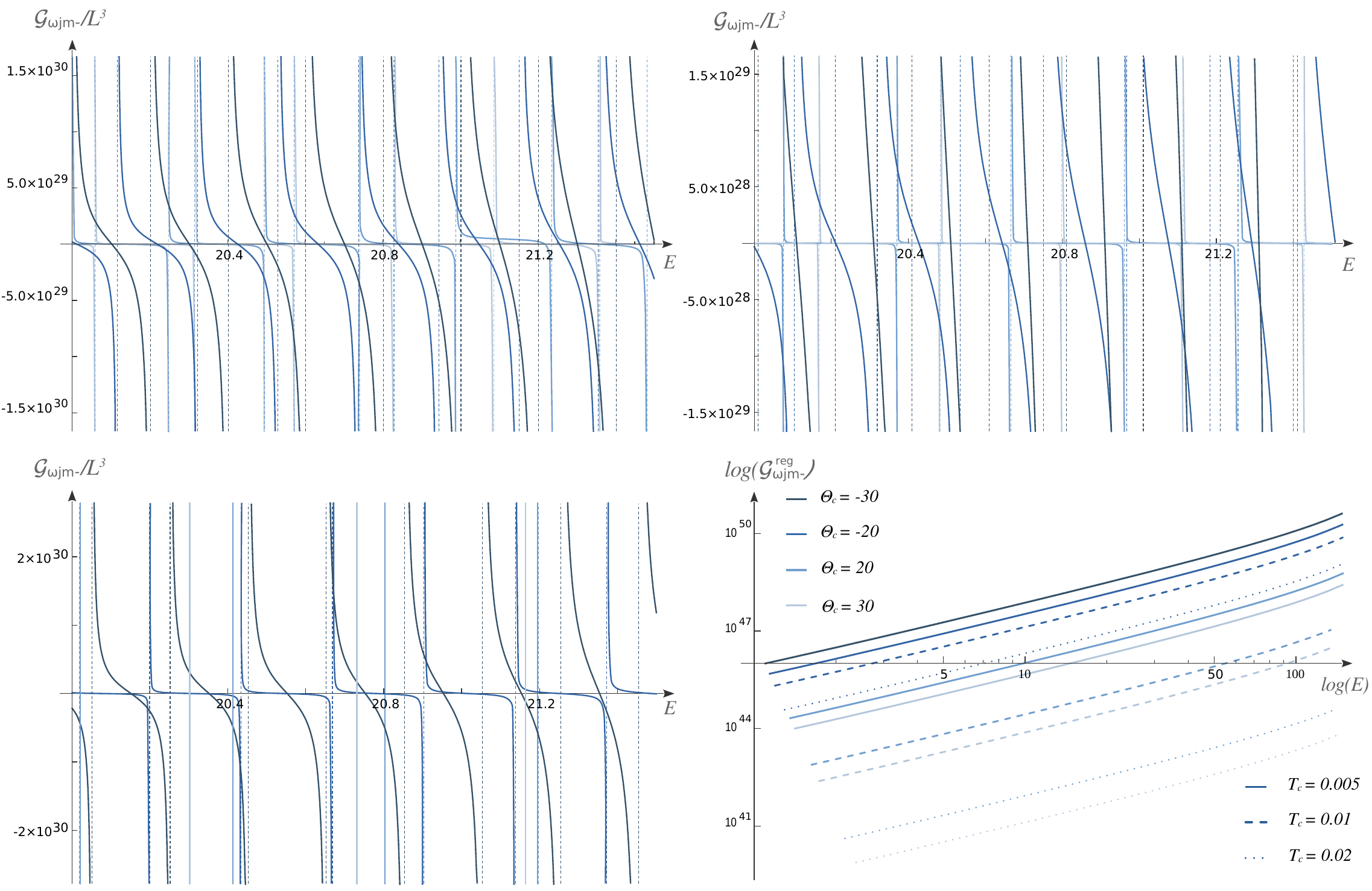}
\caption{Plots of the fermionic two-point correlator, corresponding to three different central temperatures $T_{\sf c}=0.005$ (top left), $T_{\sf c}=0.01$ (top right) and $T_{\sf c}=0.02$ (bottom left), each with four different values of the central degeneracy $\Theta_{\sf c}=-30,\,-20,\,20,\,30$. Also plots of the regularized correlator for the same values of the central temperatures and central degeneracies (bottom right). In all the plots we put ${\sf m}L=10$.}
\label{fig:correlators}
\end{figure}

\section{Results and discussion}
\label{sec:results}
We obtained the two point fermionic correlator \eqref{eq:correlators.WKB} as a function of the energy $E$, the results can be seen in the first three plots of Fig.~\ref{fig:correlators}. Notice that, for all the studied central temperatures $T_{\sf c}$, the poles get separated as we move into larger values of the central degeneracy $\Theta_{\sf c}$, entering into the unstable critical metal region.

\begin{figure}[t]
\includegraphics[width=\textwidth]{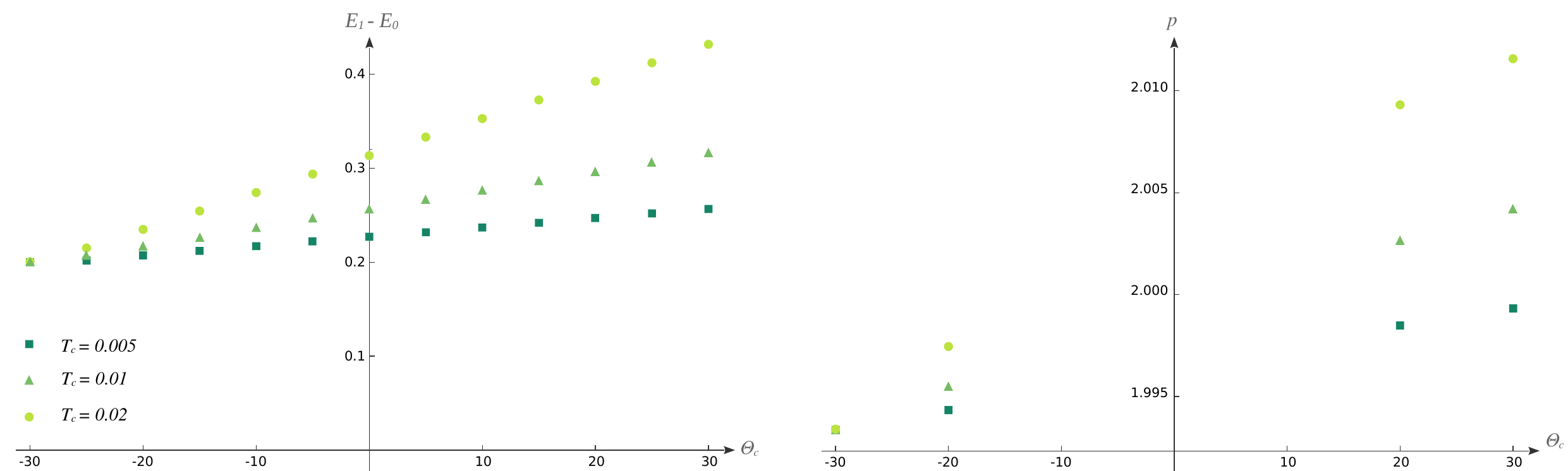}
\caption{\underline{Left:} energy distance betweed two successive poles, for different central temperatures $T_c=0.005,\,0.01,\,0.002$. The poles gets separated as the central degeneracy $\Theta_{\sf c}$ is increased, at larger temperatures they separate faster.
\underline{Right:} Power of the regularized correlator as a function of the central degeneracy for the same different central temperatures. The power grows with the temperature. In both plots we have ${\sf m}L=10$. }
\label{fig:distance.and.power}
\end{figure}

To make the aforementioned effect more evident, in Fig.~\ref{fig:distance.and.power} (left) we plotted the dependence of the energy distance between successive poles as a function of the central degeneracy $\Theta_{\sf c}$ for different values of the central temperature  $T_{\sf c}$. It is evident  that the poles separate as we increase the central degeneracy, and that they do it faster for larger temperatures.

\bigskip 
The regularized correlator \eqref{eq:regular} was also evaluated using the expression for the residua given in the Apendix \ref{app:WKB} equation \eqref{eq:residua}. The resulting plot is shown in Fig.~\ref{fig:correlators} (bottom-right) in a logarithmic scale. It is manifest in the plots that there is a power law dependence of the generalized correlator on the energy variable ${\cal G}^{\sf reg}_{EJ+}\propto E^p$ with a real power $p\in\mathbb{R}$. 

To get more information, we calculated the corresponding power $p$ for different central temperatures $T_{\sf c}$ as a function of the central degeneracy $\Theta_{\sf c}$, the result can be seen in Fig.~\ref{fig:distance.and.power} (right). The power gets larger as the central degeneracy is increased, and it does it faster at larger temperatures.

\medskip

Combining the above observations, we conclude that the dependence of the correlator on the energy has two main contributions: a pole structure and a power law regular part. As we move into the unstable critical metal region, the poles gets separated and the power on the regular part gets larger. Both effects occur faster at larger temperatures. This allows us to state that the power law contribution to the correlator gets more important as the central degeneracy grows, and effect which is more marked  at larger temperatures.

\begin{figure}[ht]
\center
\includegraphics[width=1.05\textwidth]{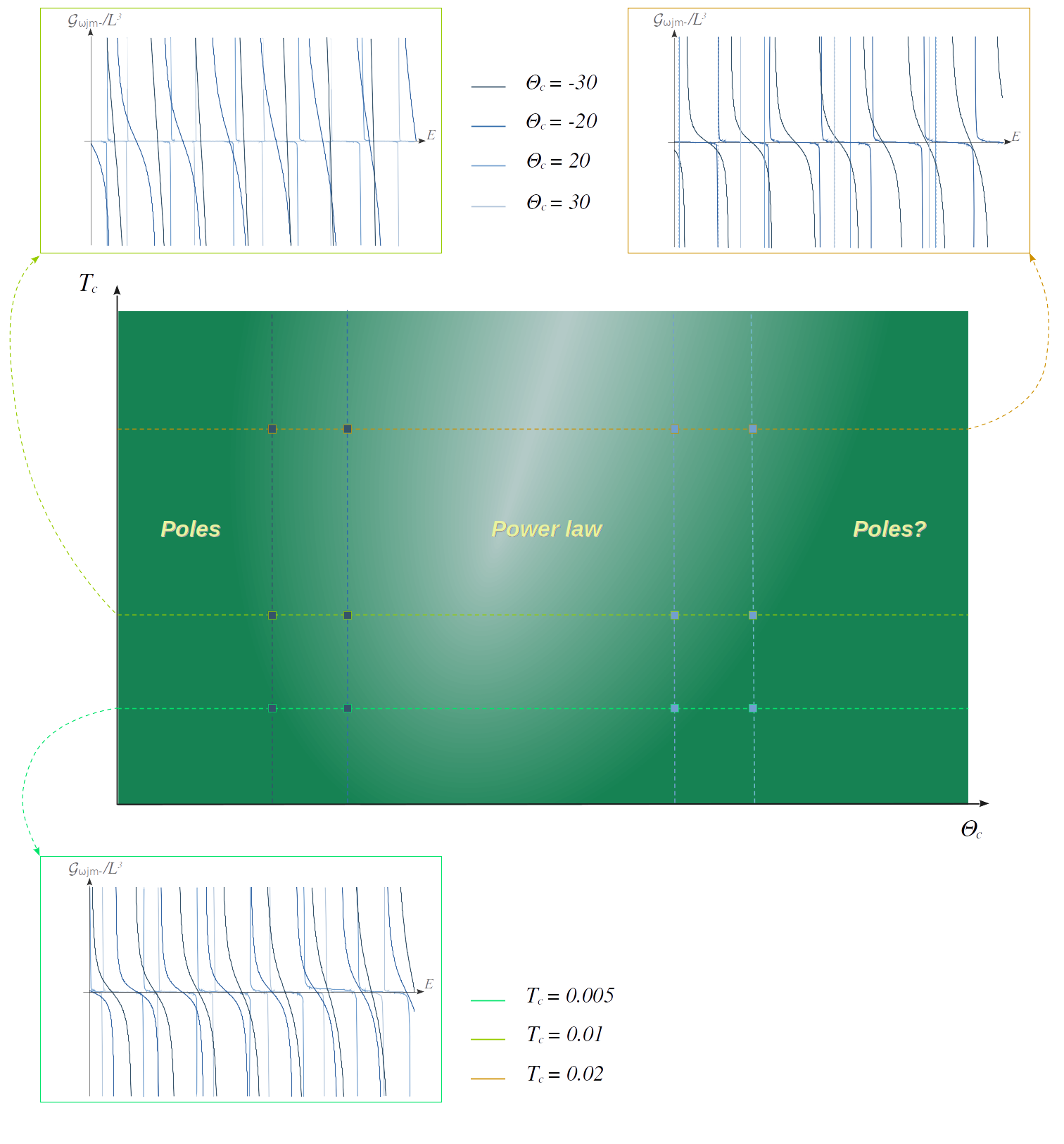}
\vspace{-.7cm}
\caption{
Phase diagram in the central degeneracy $\Theta_{\sf c}$ versus central temperature $T_{\sf c}$ plane. The poles on the fermionic two-point correlator get separated as we move into the {unstable critical metal} region, the power on its regular term becomes larger. This indicates the presence of a critical behavior.}
\label{fig:phasediagram}
\end{figure}

The above results contribute to the hypothesis of the criticality of the central region of the phase diagram of the holographic neutron star, stated in \cite{Canavesi:2023sqx}. The critical features, previously found in the density profile as a function of the radius and in the dependence of the scalar two-point correlator on the energy, are now also manifest in the fermionic counterpart. 
This is illustrated in the phase diagram of Fig. \,\ref{fig:phasediagram}.

\newpage

The conclusion is that, when finite volume effects are taken into account, the holographic description of a strongly coupled metal results in a phase diagram very similar to that of the metallic degrees of freedom of a High T$_c$ superconductor. In particular, there is a central region at intermediate dopings and low temperatures, which gets wider as the temperature is increased, where the system manifest critical features, which can be attributed to a quantum critical point. 

Since the system is unstable in the critical zone, a natural question is: What would it collapse into? Notice that it cannot become a neutral black hole, since our boundary conditions at infinity include a finite chemical potential that would then diverge at the horizon due to redshift effects. A possible answer, which we plan to explore in a forthcoming publication, is to include charged matter, \emph{i.e.} to study a holographic electron star in global AdS. 

\bigskip

\section*{Acknowledgements}

The authors are grateful to Pablo G. Pisani, Carlos Arg\"uelles, Guillermo Silva, Diego Correa, Octavio Fierro and Julio Oliva for helpful comments. 
This work is partially supported by CONICET grant PIP-2023-11220220100262CO and UNLP grant 2022-11/X931. Lucas Acito wants to thank Abdus Salam ICTP for hospitality and support during the late stages of this work.

\newpage

\appendix

\section{The holographic neutron star}
\label{app:neutronstar}

The holographic neutron star background is  a spherically symmetric metric
\bea
\label{eq:metric0}
ds^2 &=&  L^2\left(-e^{\nu(r)}\, dt^2+ e^{\lambda(r)}\,dr^2 + r^2\, d\Omega^2\right) \,,
\eea 
where $\;d\Omega^2=d\vartheta^2 + \sin^2\!\vartheta\, d\varphi^2\;$ is the metric of a two-dimensional sphere. 

The solution is sourced by a perfect gas with pressure and density obtained from the Thomas-Fermi approximation
\begin{equation}
\rho(r) = \frac{g\,GL^2}{8\pi^3}\int f(r,\mathbf{p})\sqrt{\mathbf{p}^2+{\sf m}^2 }\,d^3\mathbf{p}\,,
\quad\quad 
P(r) = \frac{g\,GL^2}{24\pi^3}\int
    f(r,\mathbf{p})\frac{\mathbf{p}^2}{\sqrt{\mathbf{p}^2+{\sf m}^2 }} \,d^3\mathbf{p}\,,~~
\label{eq:Thomas-Fermi}
\end{equation}
where  ${\sf m}$ is the mas of the fermionic gas particle, $g$ is the number of fermionic species, and the dimensional factor $GL^2$ was introduced for later convenience. Here $f(r, \mathbf{p})$ is the Fermi-Dirac distribution function
\begin{equation}
f(r,\mathbf{p})=\left(1+{\exp\left({\frac{\sqrt{\mathbf{p}^2+ {\sf m}^2}-\mu(r)}{T(r)}}\right)} \right)^{-1}\,.
\label{eq:fermi-dirac-distribution}
\end{equation}
This depends on the local temperature  $T(r)$ and chemical potential  $\mu(r)$, defined at equilibrium by Tolman and Klein conditions respectively
\begin{align}
\label{eq:Tolman.Klein}
e^{\nu(r)/2}\, T(r) = \mbox{constant}\,,
\qquad\qquad
e^{\nu(r)/2}\, \mu(r) = \mbox{constant}\,.
\end{align}

Inserting the Ansatz \eqref{eq:metric0} in the Einstein equations with a negative cosmological constant $-3/L^2$, results in the Tolman-Openheimer-Volkoff system of coupled ordinary differential equations
\begin{equation}
\label{eq:TOV}
\frac{dM}{dr}=4\pi{r}^2 \rho(r) \,.
\qquad\qquad\quad
\frac{d\chi}{dr}=8\pi r\left(P(r)+\rho(r)\right)e^{\lambda(r)}  
\end{equation}
where we have re-defined $\chi=\nu+\lambda$ and $e^{\lambda(r)}=\left(1-\frac{2 M(r)}{r}+r^2\right)^{-1}$. 

Equations \eqref{eq:TOV} have to be solved numerically by shooting, with regular boundary conditions at the centre of the star $M(0)=0$ and $\chi(0)=0$. This implies $\nu(0)=0$ and thus the constants in \eqref{eq:Tolman.Klein} correspond to the central temperature $T_{\sf c}$ and the central chemical potential $\mu_{\sf c}$ respectively. 
As the local chemical potential $\mu(r)$ becomes smaller than the particle mass ${\sf m}$, what happens for large $r$, the Fermi-Dirac distribution gets damped and the density and pressure vanish. We have reached the boundary of the star, the mass $M(r)$ becomes constant and the solution transitions smoothly into a Schwazchild-AdS metric. 
\newpage
\section{The Dirac operator on the sphere}
\label{app:DiracopS2}

The eigenvalue problem of the Dirac operator on the sphere was tackled in many references in the past. Here we give a brief review of the case of interest, namely the two-dimensional sphere $S^2$, following closely \cite{Abrikosov:2002jr}.

We start by writing the metric of the sphere in the form
\bea
d\Omega^2&=&d\vartheta^2+\sin^2\!\vartheta \,d\varphi^2=\delta_{IJ}\,\hat\omega^I\,\hat\omega^J \,,\qquad\qquad\quad I,J\in\{1,2\}\,,
\eea
where in the second equality we have introduced a local frame, {\em i.e.} a zweibein basis $\hat \omega^I$ and dual vector field $\hat e_I$, which satisfy $\hat \omega^I(\hat e_J) = \delta^I{}_J$. These objects can be written  as
\begin{align}
\hat\omega^1 =  d\vartheta\,,
\qquad\quad
\hat\omega^2  =    \sin\vartheta\, d\varphi \,, 
\qquad \quad \mbox{and}\quad \qquad \hat e_1 =  \partial_\vartheta\,,\qquad  
\quad  \hat e_2 = \frac1{ \sin\vartheta}\,\partial_\varphi\,.
\end{align}
The resulting spin connection reads 
\bea
\hat\omega^1{}_2=-\cos\vartheta\,d\varphi\,.
\eea
A suitable choice of gamma-matrices satisfying $\{\gamma^I,\gamma^J\} = 2\,\delta^{IJ}$ is
\bea
\gamma^1=\sigma_1\,,\qquad\qquad \qquad\quad 
\gamma^2=\sigma_2\,.
\eea
With this, 
%the rotation generator takes the form
%\bea
%S_{12}=\frac{1}{2}\,\gamma_1\,\gamma_2 = \frac i2{\sigma_3}
%\eea
%%
%and then 
the Dirac operator reads
\be
\hat{\slashed{\nabla}} 
= \sigma_1 \,\left(\partial_\vartheta + \frac{\cot\vartheta}{2}\right)+ \sigma_2\,\frac1{\sin\vartheta}\,\partial_\varphi\,,
\ee
resulting in the eigenvalue equation 
\be\label{eq:eigenDeq}
\hat{\slashed{\nabla}}\psi_\alpha(x,\varphi) = -i\,\alpha\;\psi_\alpha(x,\varphi)\,.
\ee
It is convenient to introduce the coordinate: $x= \cos\vartheta \in[-1,+1]$. If we now write a generic spinor of defined Fourier mode $m\in\mathbb{Z}+{1}/{2}$ as,
\be\label{eq:Psim}
\psi_{\alpha m}(x,\varphi) = \frac{e^{i\,m\,\varphi}}{\sqrt{2\pi}}\;
\left(1-x^2\right)^{-\frac{1}{4}}\;
\left(\begin{array}{c}\phi_{\alpha m}^+(x)\\
	\phi_{\alpha m}^-(x)\end{array}\right)\,,
\ee
we get the following coupled system for the components $\phi_{m}^\pm(x)$,
\be\label{eq:coupledDE}
\sqrt{1-x^2}\;\phi_{\alpha m}^\pm{}'(x) \pm \frac{m}{\sqrt{1-x^2}}\;
\phi_{\alpha m}^\pm(x) = i\,\alpha\;\phi_{\alpha m}^\mp(x)\,.
%\cr\sqrt{1-x^2}\;\phi'{}_{\lambda m}^-(x) - \frac{m}{\sqrt{1-x^2}}\;
%\phi_{\lambda m}^-(x) &=& i\,\lambda\;\phi_{\lambda m}^+(x)
\ee
If $\alpha=0$ it is straight to integrate them, obtaining singular solutions at the poles of the sphere $x=\pm 1$ (even for $m=0$ the constant solution is singular, in view of (\ref{eq:Psim})). 
So there is no eigenfunction with null eigenvector\footnote{
This is a particular case of a general result known as the Lichnerowicz theorem, which roughly states that the Dirac operator does not admit zero modes on manifolds of positive curvature  \cite{Li1980}.
}.
Then we can safely proceed to get second order differential equations, introducing in the down (respectively up) equation the value of 
$\phi_{\alpha m}^-(x)$ (respectively $\phi_{\alpha m}^+(x)$) giving by the up  (respectively down) equation,
\be\label{eq:2order1}
(1-x^2)\,\phi_{\alpha m}^\pm{}''(x) - x\,\phi_{\alpha m}^\pm{}'(x)
+\left( \alpha^2 + \frac{m(-m\pm x)}{1-x^2}\right)\,
\phi{}_{\alpha m}^\pm(x) =0\,.
\ee
As a last step, we put the above equation in a recognizable  hypergeometric form, 
through  the redefinitions
\be\label{eq:phiP}
\phi{}_{\alpha m}^\pm(x) \equiv (1+x)^{\frac{1}{4}+\frac12{|m\pm\frac{1}{2}|}}\;
(1-x)^{\frac{1}{4}+\frac12{|m\mp\frac{1}{2}|}}\;
P_{\alpha m}^\pm(x)\,,
\ee
and then by inserting (\ref{eq:phiP}) in (\ref{eq:2order1}) we get,  
\be
\label{eq:2order2}
(1-x^2)\,P_{\alpha m}^\pm{}''(x) 
+ 2\,\left(a_\pm-b_\pm\,x\right)P_{\alpha m}^\pm{}'(x)
+\left(\alpha^2- \left(b_\pm-\frac{1}{2}\right)^2\right)
P_{\alpha m}^\pm(x)=0 \,,
\ee 
where we called $a_\pm\equiv\left(|m\!\pm\!{1}/{2}|-|m\!\mp\!{1}/{2}|\right)/2$ and $b_\pm\equiv1+\left(|m\!\pm\!{1}/{2}|+|m\!\mp\!{1}/{2}|\right)/2$.
These equations have as unique regular solutions the Jacobi polynomials 
$P_{j-|m|}^{\left(|m\!\mp\!{1}/{2}|,|m\!\pm\!{1}/{2}|\right)}$ with {half-integer} index $j\geq |m|$, provided that the eigenvalue has the form
\be\label{eq:eigenvalue}
\alpha_{j\epsilon} = \epsilon\,\left(j+\frac{1}{2}\right)\,,\qquad\qquad \qquad \epsilon^2=1\,.
\ee
Summarizing, the most general regular solution (\ref{eq:Psim}) to the Dirac equation (\ref{eq:eigenDeq}) has the form
\be\label{eq:Psigral}
\psi_{jm\epsilon}(x,\varphi) = \frac{e^{i\,m\,\varphi}}{\sqrt{2\pi}}\;
\left(\begin{array}{c}
C_{jm\epsilon}^+\;(1-x)^\frac{|m-\frac{1}{2}|}{2}\;(1+x)^\frac{|m+\frac{1}{2}|}{2}\;
P_{j-|m|}^{(|m-\frac{1}{2}|,|m+\frac{1}{2}|)}(x)\\
C_{jm\epsilon}^-(1-x)^\frac{|m+\frac{1}{2}|}{2}\;(1+x)^\frac{|m-\frac{1}{2}|}{2}\;
P_{j-|m|}^{(|m+\frac{1}{2}|,|m-\frac{1}{2}|)}(x)
\end{array}\right)\,,
\ee
with the eigenvalue given in (\ref{eq:eigenvalue}).
However the constants $C_{jm\epsilon}^\pm$ are not independent, 
because to arrive to \eqref{eq:2order1} from \eqref{eq:coupledDE} 
one has to express one component in terms of the other. 
Compatibility leads to the relation, 
\be\label{eq:C-C+}
C_{jm\epsilon}^- = i\,\sign(m)\,\epsilon\;C_{jm\epsilon}^+\,.
\ee
The solution is then determined up to un overall constant $C_{jm\epsilon}^+$. 
Its modulus can be fixed by imposing normalization with respect to the usual scalar product
\be
(\psi_{jm\epsilon};\psi_{j'm'\epsilon'})= 
\int_{-1}^{+1}dx\;\int_{0}^{2\pi}d\varphi\;
\psi^\dagger_{jm\epsilon}(x,\varphi)\psi_{j'm'\epsilon'}(x,\varphi) 
\equiv \delta_{jj'}\;\delta_{mm'}\;\delta_{\epsilon\epsilon'}\,,
\ee
while the phase can be fixed by imposing the complex conjugation rule
\be 
-i\,\sigma_2\;\psi_{jm\epsilon}^* 
= (-1)^{j+m}\;\psi_{j(-m)\epsilon}\,.
\ee

%As a final remark, we note that the solutions at fixed $\epsilon$, i.e.  positive or negative eigenvalues $\pm(j+\frac{1}{2})$, are encoded in the half-integer representations  of the isometry group $SU(2)$. This is so because there exist operators $J_I$ such that $[\slashed\nabla; J_I] =0$. But there also exists a parity operator $P:\{P;\gamma_I\}=0$ ($P=\sigma_3$ in our conventions), such that $[P; J_I] =0$.  So we can have irreducible representations of $SU(2)$ of definite parity. Indeed, these are the up and down components of (\ref{eq:Psigral}). However $\slashed\nabla$ and $P$ does not commute, instead $\{\slashed\nabla; P\}=0$. The action of $P$ is then changing the sign of  $\lambda_{j\epsilon}$, (if you like, act with $P$ on (\ref{eq:Psigral}) and recall (\ref{eq:C-C+})). Thus the doubling by parity is at the origin of the existence of  positive as well as negative eigenvalues in the same representation. See \cite{Abrikosov:2002jr} for details.

\newpage
\section{The Dirac operator on the neutron star}
\label{app:DiracopAdS}

Let us now consider a $3\!+\!1$-dimensional space-time endowed with a metric of the form defined in \eqref{eq:metric0}
\bea
\label{eq:metric}
ds^2 &=&  L^2\left(-e^{\nu(r)}\, dt^2+ e^{\lambda(r)}\,dr^2 + r^2\, d\Omega^2\right)=\eta_{AB}\,\omega^A\,\omega^B\,, \quad\  A,B\in\{0,1,2,3\}\,.~~~~
\eea 
Again in the second line we have introduced a local frame, with a vielbein basis $\omega^A$ and dual vector field $e_A$ satisfying $\omega^A(e_B) = \delta^A{}_B$. Explicitly it reads
\begin{align}
\omega^0 &=  L\,e^{\frac{\nu(r)}2}\;dt\,,
&\ \
\omega^1 &= L\,r\,d\vartheta\,,\qquad
&~~&
e_0 &= \frac1L\,e^{-\frac{\nu(r)}2}\;\partial_t \,,
&
&\ \ \
e_1 &= \frac{1}{L\,r}\;\partial_\vartheta\,,
\nonumber\\
\omega^2 &= L\, r\,\sin\vartheta\, d\varphi \,,
&\ \
\omega^3 &= L\,e^{\frac{\lambda(r)}2}\,dr\,,
&~&
e_2 &= \frac1{L\,r\,\sin\vartheta}\,\partial_\varphi\,,
&&\  \ \
e_3 &= \frac1L\,e^{-\frac{\lambda(r)}2}\,\partial_r\,.~~
\end{align}
The one-form connections for this geometry result
\begin{align}
&\omega_{01}=0\,,&&\ \
\omega_{02}=0\,,&&\ \
\omega_{03} = -\frac{1}{2}\,e^{\frac{\nu(r)-\lambda(r)}2}\,\nu'(r)\,dt\,,&&
\\
&\omega_{12}= - \cos\vartheta\,d\varphi \,,
&& \ \
\omega_{13}= e^{-\frac{\lambda(r)}{2}}\,d\vartheta \,,
&&\ \
\omega_{23}= e^{-\frac{\lambda(r)}{2}}\,\sin\vartheta\,d\varphi\,.
\end{align}
Since space-time is naturally divided in a $(t, r)$-spacetime and a two-dimensional sphere, it is convenient to choose the local gamma-matrices obeying $\{\Gamma^A,\Gamma^B\} = 2\,\eta^{AB}$ as follows
\be\label{eq:gamma-matrices}
\Gamma^0\equiv i\,\sigma_1\otimes 1_{2\times 2}\,,
\qquad \qquad
\Gamma^I\equiv\sigma_2\otimes\gamma^I\,,\qquad \qquad
\Gamma^3\equiv\sigma_3\otimes 1_{2\times 2}\,,
\ee
where $\{\sigma_i, i=1,2,3\}$ are the Pauli matrices and 
$\gamma^I$ correspond to the gamma matrices on the sphere as defined in Appendix \ref{app:DiracopS2}.
This choice of representation leads to real fermion equations, see below. 
 
In the above defined background we consider the Dirac equation for fermions of mass ${\sf m}$ described by a four-component Dirac spinor $\Psi$, writing 
\be\label{eq:eom-fermionic}
(\slashed\nabla -i\,\slashed A - {\sf m})\,\Psi(t,r,\vartheta,\varphi) =0\,,
\ee
where $\slashed\nabla\equiv \Gamma^A\nabla_A= 
\Gamma^A\left(e_A +{\omega_A{}^{BC}}[\Gamma_B;\Gamma_C]/8\right)$ is the covariant derivative, and 
$\slashed A =\Gamma^B \,A_B\;$ accounts for the coupling to a gauge potential. We will define this last term as a constant chemical potential $A_t=\mu$, resulting in $\slashed A=({\mu}/{L})\, e^{-\frac {\nu(r)}2}\,\Gamma^0 $.
 
Taking into account the decomposition (\ref{eq:gamma-matrices}), we consider a complete basis $\psi_{jm\epsilon}$ of spinors on the sphere, which are eigenfunctions of the Dirac operator explained in Appendix \ref{app:DiracopS2}. In terms of them, we can write 
\be
\label{eq:Psiexpan}
\Psi(t,r,\vartheta,\varphi) = 
\frac{e^{-\frac{\nu(r)}{4}}}{r}\,
\sum_{jm\epsilon} \int\frac{d\omega}{2\pi}\;
\varphi_{\omega  jm\epsilon}(r)\otimes \psi_{jm\epsilon}(\vartheta,\varphi)\,e^{-i\,\omega t}\,.
\ee 
By plugging (\ref{eq:Psiexpan}) in \eqref{eq:eom-fermionic} we get for the bi-spinor 
$\varphi_{\omega \,jm\epsilon}(r)$ the equation 
\be\label{eq:fieq1}
\varphi'_{\omega  jm\epsilon}(r)+ e^{\frac{\lambda(r)}2}\;\left(
i\,(\omega+\mu)\,e^{-\frac{\nu(r)}2} \,\sigma_2+ \frac{\alpha_{j\epsilon}}{r}\; \sigma_1 - {\sf m}L\;\sigma_3\right)\; \varphi_{\omega  jm\epsilon}(r) =0\,.
\ee
In terms of the components $\varphi_{\omega  jm\epsilon}=\left(\begin{array}{c}\varphi^{+}_{\omega  jm\epsilon}\\\varphi^{-}_{\omega  jm\epsilon}\end{array}\right)$ we have the coupled equations 
\bea
\label{eq:fieq2}
\varphi^{+'}_{\omega  jm\epsilon}(r) 
- {\sf m}L\,
e^{\frac{\lambda(r)}{2}}
\,\varphi^+_{\omega  jm\epsilon}(r)  
+
e^{\frac{\lambda(r)}2} 
\left(({\omega+\mu})\,e^{-\frac{\nu(r)}{2}} +\frac{\alpha_{j\epsilon}}{ r}\right) 
\varphi_{\omega  jm\epsilon}^-(r)&=&0\,,
\\ 
\label{eq:fieq22}
\varphi^{-'}_{\omega  jm\epsilon}(r) 
+{\sf m}L\, e^{\frac{\lambda(r)}{2}}\varphi^-_{\omega  jm\epsilon}(r)  
-e^{\frac{\lambda(r)}2}\left(({\omega+\mu})\,e^{-\frac{\nu(r)}{2}}-\frac{\alpha_{j\epsilon}}{ r}\right) 
\varphi_{\omega  jm\epsilon}^+(r)&=&0\,. \ \ ~
\eea
If $\epsilon={\rm sgn}(\alpha_{j\epsilon})= + 1$ (respectively $-1$), from the first (respectively the second) equation in (\ref{eq:fieq2}) 
we can write $\varphi^-_{\omega jm\epsilon}(r)$ (respectively $\varphi^+_{\omega jm\epsilon}(r)$) in terms of 
$\varphi^+_{\omega jm\epsilon}(r)$ (respectively $\varphi^-_{\omega jm\epsilon}(r)$), resulting in 
\be\label{eq:varphi-ee}
\varphi^{-\epsilon}_{\omega jm\epsilon}(r) = \frac{1}{f_{\omega  j }(r)} 
\left(-\epsilon\,\varphi'^{\epsilon}_{\omega  jm\epsilon}(r) +  
{\sf m}L\;e^{\frac{\lambda(r)}{2}}\;\varphi^\epsilon_{\omega jm\epsilon}(r)\right)\,,
\ee
where we have defined the positive function,
\be\label{eq:fomegaj}
f_{\omega  j }(r)=e^{\frac{\lambda(r)}{2}}\;\left({(\omega+\mu)}e^{-\frac{\nu(r)}{2}} + \frac{|\alpha_{j\epsilon}|}{ r}\right)\,.
\ee
By plugging $\varphi^-_{\omega jm\epsilon}(r)$ (respectively $\varphi^+_{\omega jm\epsilon}(r)$)  
in the second (respectively first) equation in (\ref{eq:fieq2}), and after a further rescaling
$
\varphi^{\epsilon}_{\omega jm\epsilon}(r) \equiv \sqrt{f_{\omega  j }(r)}\;\phi_{\omega jm\epsilon}(r)
$,
we see that 
%{\color{blue}
%\be
%\label{eq:varphi-ee2}
%\varphi^{-\epsilon}_{\omega jm\epsilon}(r) = \frac{-\epsilon}{\sqrt{f_{\omega  j }(r)}} 
%\left(\phi'_{\omega  jm\epsilon}(r) 
%+
%\left(\frac{f'_{\omega  j }(r)}{2{f_{\omega  j }(r)}}-\epsilon{\sf m}L\;e^{\frac{\lambda(r)}{2}}\right)\phi_{\omega  jm\epsilon}(r)  
%\right)
%\ee}
%
%
the functions $\phi_{\omega \,jm\epsilon}(r)$ satisfy the Schr\"odinger  equations
\be\label{eq:Schroeq}
\phi''_{\omega  jm\epsilon }(r) - U_{\omega  j \epsilon }(r)\;\phi_{\omega jm\epsilon }(r)=0\,,
\ee
with potentials
\small
\bea\label{eq:Upot}
U_{\omega j \epsilon }(r) &=&\!
\sqrt{f_{\omega  j  }(r)}\left(\frac{1}{\sqrt{f_{\omega  j  }(r)}}\right)'' \!\!\!
+\epsilon  {\sf m}L\,f_{\omega j}(r) \left(\frac{e^{\frac {\lambda(r)} 2}}{f_{\omega  j  }(r)}\right)'\!\!\!
%\cr&-& 
-e^{\lambda(r)}\!
\left(\frac{(\omega+\mu)^2}{e^{\nu(r)}} \!-\! \frac{\alpha_{j\epsilon  }{}^2}{r^2} \!-\! {\sf m}^2 L^2\right)\,.\nonumber\\
\eea
\normalsize
The spinor $\varphi_{\omega jm\epsilon}$ can then be written as,
\small
\bea\label{eq:spinorphi2}
\varphi_{\omega  jm\epsilon}(r)\!=\!  \sqrt{f_{\omega  j}(r)}
\left(
\begin{array}{c}
\frac{1+\epsilon}2
\phi_{\omega  jm+}(r)
+ 
\frac{1-\epsilon}{2f_{\omega j}(r)}\!
\left(\phi'_{\omega  jm-}\!(r) + \left(\frac{f'_{\omega j}(r)}{2\,f_{\omega j}(r)}
+{\sf m} L\,e^{\frac{\lambda(r)}{2}}\right)\phi_{\omega  jm-}(r)\right)
\\ ~ \\
\frac{1-\epsilon}2\phi_{\omega  jm-}(r)
-\frac{{1+\epsilon}}{2f_{\omega  j }(r)}\! 
\left(\phi'_{\omega  jm+}\!(r) + \left(\frac{f'_{\omega  j }(r)}{2\,f_{\omega  j }(r)}
-{\sf m}L\,e^{\frac{\lambda(r)}{2}}\right)\phi_{\omega  jm+}(r)\right)
\end{array}\!
\right)  \nonumber 
\eea
\be\ee
\normalsize

\newpage

\section{Fermionic correlators in global AdS}
\label{app:correlator}

An action for the Dirac equation \eqref{eq:eom-fermionic} can be written in terms of the spinor $\Psi(t,r,\vartheta,\varphi)$ as
\be
\label{eq:action-fermionic}
S_{\sf Dirac} = -\int d^4x\sqrt{|g|}\;\bar\Psi \left(\slashed D - {\sf m} \right) \Psi
- \int_{r\rightarrow\infty} d^3 x\,\sqrt{|h|}\; \bar\Psi_+ \Psi_-\,,
\ee
where  the conjugate spinor  is  $\;\bar\Psi\equiv i\Psi^{\dagger}\Gamma^0$, and the projections $\Psi_\pm $ are defined according to $\Psi_\pm \equiv \frac12(1\pm \Gamma^3)\Psi$. 
In the boundary term $h_{\mu\nu}$ is the induced metric. This term is needed to have a well-defined variational principle and it depends on the boundary conditions to be imposed \cite{Henningson:1998cd}. 
The form chosen in \eqref{eq:action-fermionic} corresponds (for ${\sf m}>0$) to fix the right-handed fermion $\Psi_+$ as the source on the boundary.%\footnote{ If we allow for a negative mass, the boundary term must be $+\int_{r\rightarrow\infty} d^d x\,\sqrt{|h|}\; \bar\Psi_-\;\Psi_+$, that corresponds to fix the left-handed component $\Psi_-$ of the spinor field on the boundary. These fact is related to the near boundary expansion \eqref{eq:gralUV} where, to get a smooth solution in absence of the source, the source term must correspond to the leading order term \cite{Mueck:1998wkz}. When ${\sf m}=0$ we can chose any of the two posibilities.}.

In the Schr\"odinger equations \eqref{eq:Schroeq} the potential near the boundary takes the form $U_{\omega j\epsilon}(r)\sim {\sf m}L({\sf m}L+\epsilon)/r^2$, implying for the asymptotic form of the solution
\be
\label{eq:culoculo}
\phi_{\omega jm\epsilon} (r)
\approx
A_{\omega jm\epsilon} \left(r^{1+\epsilon {\sf m}L}+\dots\right)+
B_{\omega jm\epsilon} \left(r^{-\epsilon{\sf m}L}+\dots\right)\,,
\ee
for some constants $A_{\omega jm\epsilon}$ and $ B_{\omega jm\epsilon}$. This results in the asymptotic  form of the spinor
% {\color{blue}
%\be
%\label{eq:varphi-ee22}
%\varphi^{\epsilon}_{\omega jm\epsilon}(r) = \sqrt{\omega+\mu+|\alpha_{j\epsilon}|}\,
%\left(A_{\omega jm\epsilon} \left(r^{\epsilon {\sf m}L}+\dots\right)+
%B_{\omega jm\epsilon} \left(r^{-\epsilon{\sf m}L-1}+\dots\right)
%\right)
%\ee}
%{\color{blue}
%\be
%\label{eq:varphi-ee2}
%\varphi^{-\epsilon}_{\omega jm\epsilon}(r) = \frac{1}{\sqrt{\omega+\mu+|\alpha_{j\epsilon}|}} 
%\left(
%\epsilon+2{\sf m}L
%\right)B_{\omega jm\epsilon} \left(r^{-\epsilon{\sf m}L}+\dots\right)
%\ee}
%{
%\color{green}
%\be\label{eq:varphi-ee111}
%\varphi^{-\epsilon}_{\omega jm\epsilon}(r) = \frac{1}{\sqrt{\omega+\mu+|\alpha_{j\epsilon}|}} 
%\left(2{\sf m}L+\epsilon\right)B_{\omega jm\epsilon} \left(r^{-\epsilon{\sf m}L }
%+\dots\right)\ee
%}
%
%$$ 
%A_{\omega jm\epsilon} 
%\frac{1-\epsilon}{2}
%r^{ {\sf m}L}(2{\sf m}L-1)
%+
%B_{\omega jm\epsilon} r^{ -{\sf m}L}
%\frac{1+\epsilon}2(1+2{\sf m}L) 
%$$
%	
\bea\label{eq:spinorphi}
\varphi_{\omega  jm\epsilon}(r)\approx  {\cal J}_{\omega j m \epsilon}\,r^{ {\sf m}L}\left(
\left(
\begin{array}{c}1
\\ 0
\end{array}
\right)
+\dots
\right)
+
\frac1{L^3}\,\langle{\cal O}\rangle_{\omega j m \epsilon}\,r^{-{\sf m}L}
\left(
\left(
\begin{array}{c}0
\\ 1
\end{array}
\right)
+\dots 
\right)\,,
\eea 
in terms of new constants ${\cal J}_{\omega jm\epsilon}$ and $\langle{\cal O}\rangle_{\omega jm\epsilon}$ which can be written as
\be
{\cal 
J}_{\omega jm\epsilon}
=%\left(
\frac{1+\epsilon}2
{\sqrt{\omega+\mu+|\alpha_{j\epsilon}|}}\,
A_{\omega jm +}   
+ 
\frac{1-\epsilon}2
\frac{2{\sf m}L -1
}{\sqrt{\omega+\mu+|\alpha_{j\epsilon}|}}\,
B_{\omega jm-}
%\right)
\,,
\label{eq:sarara1}
\ee
\be
\langle{\cal O}\rangle_{\omega jm\epsilon}/L^3
=
%\left(
\frac{1-\epsilon}2 
{\sqrt{\omega+\mu+|\alpha_{j\epsilon}|}}\,
A_{\omega jm -}  
+\frac{1+\epsilon}2\frac{2{\sf m}L+1}
{\sqrt{\omega+\mu+|\alpha_{j\epsilon}|}}
\,B_{\omega jm+}
%\right)
\,.
\label{eq:sarara2}
\ee

%This behavior yields for the spinor \eqref{eq:spinorphi} 
%\be\label{eq:phiasymptotics}
%\phi_{\omega J}(r)\quad\stackrel{r\rightarrow\infty}{\longrightarrow}\quad
%\frac{\tilde\psi_{\omega J}^+}{f^\infty_{\omega J}
%{}^\frac{1}{2}\,e^{-\frac{\chi_\infty}{4}}}\;r^{1+\gamma}\;
%\left(\begin{array}{c}1\\0\end{array}\right)
%+\dots+
%\frac{\tilde\psi_{\omega J}^-}{f^\infty_{\omega J}
%{}^\frac{1}{2}\,e^{-\frac{\chi_\infty}{4}}}\;r^{1-\gamma}\;
%\left(\begin{array}{c}0\\1\end{array}\right)
%\ee
By plugging the asymptotics \eqref{eq:spinorphi} for the AdS spinor in the decomposition \eqref{eq:Psiexpan} of the full solution, we get 
\be\label{eq:Psiexpanasymp}
\Psi(t,r,\vartheta,\varphi) \approx r^{-\frac{3}{2}+{\sf m}L}\,\left( {\cal J}(t,\vartheta,\varphi)\,+\dots\right)
+ r^{-\frac{3}{2}-{\sf m}L}\, \left(\langle{\cal O}(t,\vartheta,\varphi)\rangle +\dots\right)\,,
\ee
where the right and left spinors with respect to $\Gamma^3$ are   given by
\bea
{\cal J}(t,\vartheta,\varphi) &=& 
\sum_{jm\epsilon} \int\frac{d\omega}{2\pi}\,e^{-i\,\omega t}\,
%e^{-\frac{\chi_\infty}{4}}\,
{\cal J}_{\omega jm\epsilon}%\,
\left(\begin{array}{c}1\\0\end{array}\right)
\otimes \psi_{jm\epsilon}(\vartheta,\varphi)\,,
\label{sandanga1}
\\
\langle{\cal O}(t,\vartheta,\varphi)\rangle &=& \frac{1}{L^3}
\sum_{jm\epsilon} \int\frac{d\omega}{2\pi}\,e^{-i\,\omega t}\,
%e^{-\frac{\chi_\infty}{4}}\,
\langle{\cal O}\rangle_{\omega jm\epsilon}%\,
\left(\begin{array}{c}0\\1\end{array}\right)
\otimes \psi_{jm\epsilon}(\vartheta,\varphi)\,.
\label{sandanga2}
\eea

In a holographic description the spinor ${\cal J}$ is identified with the source of a fermionic operator ${\cal O}$ at the boundary. The corresponding generating functional for connected correlators is obtained by evaluating the bulk action \eqref{eq:action-fermionic} on shell, and with the help of \eqref{eq:Psiexpanasymp}, \eqref{sandanga1} and \eqref{sandanga2}, as
\bea\label{eq:adscft2}
G\left[\bar {\cal J}, {\cal J}\right]=  \sum_{jm\epsilon} \int\frac{d\omega}{2\pi}\;
{\cal J}_{\omega jm\epsilon}^*  \langle{\cal O}\rangle_{\omega jm\epsilon}\,.
\eea
By taking the derivative of $G\left[\bar {\cal J}, {\cal J}\right]$ with respect to the source ${\cal J}(t,\vartheta,\varphi)$ we obtain the expectation value $\langle{\cal O}\rangle$ which, as our notation   suggested, is given by the spinor $\langle{\cal O}(t,\vartheta,\varphi)\rangle$. One further derivative with respect to the source ${\cal J}(t,\vartheta,\varphi)$ gives us the connected two-point correlation function. In Fourier space, that would imply to take the derivative of $\langle{\cal O}\rangle_{\omega jm\epsilon}$ with respect to ${\cal J}_{\omega jm\epsilon}$. Since it is evident from the equations \eqref{eq:sarara1}-\eqref{eq:sarara2} that they are linearly related, such derivative can be obtained simply by taking the quotient ${\cal G}_{\omega jm\epsilon}= {\langle{\cal O}\rangle_{\omega jm\epsilon}}/{{\cal J}_{\omega jm\epsilon}}$. This yields
\be\label{eq:propagators}
{\cal G}_{\omega jm+} = L^3\,\frac{{2\sf m}L +1}{\omega+\mu+|\alpha_{j\epsilon}|}\,
\frac{B_{\omega jm+}}{A_{\omega jm+}}\,,
\qquad \qquad
{\cal G}_{\omega jm-} = L^3\,\frac{\omega+\mu+|\alpha_{j\epsilon}|}{{2\sf m}L -1}\,
\frac{A_{\omega jm-}}{B_{\omega jm-}}\,.
\ee 
This implies that once we solved the Schr\"odinger equation \eqref{eq:Schroeq}, the coefficients $A_{\omega jm\epsilon}$ and $B_{\omega jm\epsilon}$ are obtained, and the connected two-point correlators can be evaluated. In what follows we solve the equation in the WKB limit. 
\newpage

\section{The WKB approximation}
\label{app:WKB}

Here we recall the basics of the WKB approximation \cite{merzbacher1998quantum} for the Schr\"odinger equation 
\be\label{eq:Schroedinger.equation}
-\phi''(r) + U(r)\;\phi(r) = 0\,.
\ee
We assume that the potential is repulsive at both boundaries, implying that for $r$ close enough to $r=0$ we have $U(r)>0$ and $U'(r)<0$. As we move into positive values of $r$ we have a first turning point $r_1$ at which $U(r_1)=0$ and $U'(r_1)<0$. At larger values of $r$ a second turning point $r_2$ appear at which $U(r_2)=0$.

Calling $r_j$ with $j\in\{1,2\}$ to the two turning points, we define the functions
\be\label{eq:WKB.waves}
u_\pm (r; r_j) = |U(r)|^{-\frac{1}{4}}\;e^{\pm \int_{r_j}^r dr'\,\sqrt{U(r')}}\,.
\ee
As can be checked by direct substitution, they are good approximate solutions as long as we are away from the turning points and then $U(r)$ is large enough. 
%
%Close to any turning point we can linearize the potential and write the solution in terms of Airy functions $A\mbox{\small $i$\normalsize}$ and $B\mbox{\small $i$\normalsize}$.
%
Then the WKB approximation for the solution at the left and the right of $r_j$ reads
\bea\label{eq:WKB.regions}
\phi^{\sf WKB}(r) &=& 
\left\{\begin{array}{lcrcl} 
	L^{(j)}_+\; u_+(r; r_j) + L^{(j)}_-\; u_-(r; r_j)\,,\;\;&~&\;\; 
	r-r_j&\ll& |U'(r_j)|^{-\frac{1}{3}}\,,\vspace{2mm}
%	\\ 
%	A^{(j)}_+\;A\mbox{\small $i$\normalsize}(w) + A^{(j)}_-\;B\mbox{\small $i$\normalsize}(w)
%	\big|_{w=\frac{r-r_i}{U'(r_j)^\frac{1}{3}}}\,, &~~~&\,
%	|r-r_j|&\ll &\left|\frac{2\,U'(r_j)}{U''(r_j)}\right|\,,\vspace{1mm}
\\ 
	R^{(j)}_+\; u_+(r; r_j) + R^{(j)}_-\; u_-(r; r_j)\,,\;\;&~&\;\; r-r_j&\gg&|U'(r_j)|^{-\frac{1}{3}}\,,
\end{array}\right. 
%&&
\eea
where $L_\pm^{(j)}, R_\pm^{(j)}$ %and $A_\pm^{(j)}$ are numerical constants to be determined in order to ensure the boundary conditions and the continuity around each turning point. 
are linearly related through the {\em connection formulas}
%The coefficients  at left and right of a given turning point are related linearly 
%
\be\label{eq:WKB.matrix}
\left(\begin{array}{c}R^{(1)}_+\\R^{(1)}_-\end{array}\right) = 
e^{-i\frac{\pi}{4}}\left(\begin{array}{cc}
	1&\;\frac{i}{2}\\\;i&\frac{1}{2}
\end{array}\right)
\left(\begin{array}{c}L^{(1)}_+\\L^{(1)}_-\end{array}\right) ,
\quad\quad\ \
\left(\begin{array}{c}R^{(2)}_+\\R^{(2)}_-\end{array}\right) = 
e^{+i\frac{\pi}{4}} \left(\begin{array}{cc}
	1&-i\\-\frac{i}{2}&\frac{1}{2}
\end{array}\right)
\left(\begin{array}{c}L^{(2)}_+\\L^{(2)}_-\end{array}\right) ,
\ee
%%
%where the matrix ${\bf M}(z_r)$ is obtained by   using the intermediate Airy form of the solution. It has the for ${\bf M}(r_2)={\bf M}$ and ${\bf M}(r_1)={\bf M^{\dagger}}$ where
%%
%\be\label{eq:WKB.matrix.explicit}
%{\bf M} \equiv e^{+i\frac{\pi}{4}}\;\left(\begin{array}{cc}
%	1&-i\\-\frac{i}{2}&\frac{1}{2}
%\end{array}\right)\,.
%\ee
%
The approximate WKB solutions around the two turning points are related by shifting the limits of integration in \eqref{eq:WKB.waves} from $r_1$ to $r_2$, 
\be
\label{eq:WKB.shifts}
u_\pm (r;r_1) = e^{\pm\int_{r_1}^{r_{2}} dr'\,\sqrt{U(r')}}\;
u_\pm (r;r_{2})\,.
\ee
Compatibility with (\ref{eq:WKB.regions}) thus implies 
\be
\label{eq:WKB.matrix.shifts}
\left(\begin{array}{c}L^{(2)}_+\\L^{(2)}_-\end{array}\right) =
\left(\begin{array}{cc}
	e^{+\int_{r_1}^{r_{2}} dr'\,\sqrt{U(r')}}&0\\
	0&e^{-\int_{r_1}^{r_{2}} dr'\,\sqrt{U(r')}}
\end{array}\right)
\left(\begin{array}{c}R^{(1)}_+\\R^{(1)}_-\end{array}\right) \,,
\ee
%%
%in terms of the matrix
%%
%\be
%{\bf W}=
%\left(\begin{array}{cc}
%	\exp{\left(+\int_{r_1}^{r_{2}} dr'\,\sqrt{U(r')}\right)}&0\\
%	0&\exp{\left(-\int_{r_1}^{r_{2}} dr'\,\sqrt{U(r')}\right)}
%\end{array}\right)\,.
%\label{eq:WKB.matrix.shitfs.explicit}
%\ee
With all the above, we can express 
%the whole set of coefficients of the solution around each of the turning points, in terms of $L^{(0)}_+$ and $L^{(0)}_-$, as follows
%
%\bea\label{eq:WKB.matrix.complete}\left(\begin{array}{c}L^{(r)}_+\\L^{(r)}_-\end{array}\right) &=& {\bf W}(z_{r-1})\;{\bf M}(z_{r-1})\;\dots{\bf W}(z_0)\;{\bf M}(z_0)\;\left(\begin{array}{c}L^{(0)}_+\\L^{(0)}_-\end{array}\right)\,,
%\nonumber\\\nonumber~\\
%\left(\begin{array}{c}R^{(r-1)}_+\\R^{(r-1)}_-\end{array}\right) &=& {\bf M}(z_{r-1})\; {\bf W}(z_{r-2})\;{\bf M}(z_{r-2})\;\dots{\bf W}(z_{0})\;{\bf M}(z_0)\;\left(\begin{array}{c}L^{(0)}_+\\L^{(0)}_-\end{array}\right)\,.\cr
%&&
%\eea
%
%From this, we can write 
the relation between the coefficients 
\bea
\label{eq:WKB.matrix.V}
\left(\begin{array}{c}R^{(2)}_+\\R^{(2)}_-\end{array}\right) 
%&=& 
%{\bf M}( {r_2}) 
%{\bf W}
% \,{\bf M}(r_1)  
%\left(\begin{array}{c}L^{(1)}_+\\L^{(1)}_-
%\end{array}\right)=
%\nonumber\\
&=&
\left(\begin{array}{cc}
	2 \cos\left(\int_{r_0}^{r_1} dr'\,\sqrt{-U(r')}\right)\ &\ 
	-\sin\left(\int_{r_0}^{r_1} dr'\,\sqrt{-U(r')}\right)\vspace{2mm}\\ 
	\sin\left(\int_{r_0}^{r_1} dr'\,\sqrt{-U(r')}\right)\ &\ 
	\frac{1}{2}\cos\left(\int_{r_0}^{r_1} dr'\,\sqrt{-U(r')}\right)
\end{array}\right)
\left(\begin{array}{c}L^{(1)}_+\\L^{(1)}_-\end{array}\right)\,.~~~
\eea
% 
%Bohr-Sommerfeld  quantization relations arise when we impose the boundary conditions  at both extremes, see below.  
 
%\bigskip

\begin{figure}[t]
\includegraphics[width=0.95\textwidth]{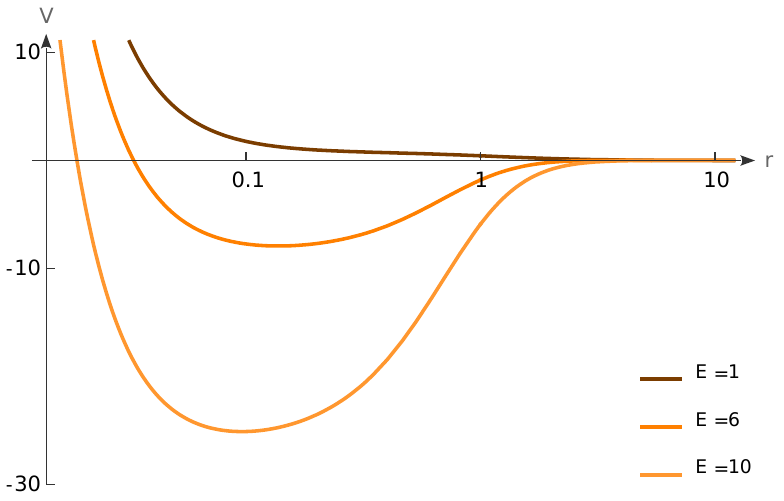}
\label{fig:potential}
\caption{Some examples of the potential of the effective Schroedinger equation, in the WKB limit. We see that, according to the value of the parameters, the potential may have two or none turning points. The plots correspond to $T_{\sf c}=0.02$, $\Theta_{\sf c}=20$ and $J=1/2$. The scale of the horizontal axis is logarithmic.}
\end{figure}

The WKB approximation is valid when the potential is large enough. For the particular case of the potential Eq. \eqref{eq:Upot} this is realized in the large ${\sf m}L$ limit, or
\be\label{eq:WKBpot2}
\left. 
U^\pm_{\omega j\epsilon}(r)\right|_{{\sf m}L\rightarrow \infty}
={\sf m}^2L^2 \; e^{\lambda(r)} \left(1+\frac{J^2}{r^2}-\frac{E^2}{e^{\nu(r)}}\right) 
\equiv
{\sf m}^2L^2\;V_{E J}(r) \,.
\ee
Here we kept fixed the scaled parameters $J\equiv(j+1/2)/{\sf m}L$ and ${E}\equiv \omega/{\sf m}L$, which are now arbitrary positive real numbers. The behaviour of the metric at the boundaries results in the asymptotics
\be\label{eq:behaV}
\left.V_{EJ}(r)\right|_{r\to \infty}
=\frac{1}{r^2}\,,
\qquad\qquad\quad
\left.V_{EJ}(r)\right|_{r\to 0}
=\frac{J^2}{r^2}\,.
\ee
This implies that the potential presents an even number of turning points. 
Explicit analysis shows that, depending on parameters, it can have two or none (see Fig.\ \ref{fig:potential}. 
Here we restrict to the case of two turning points $r_1$ and $r_2$, thus the WKB wavefunction is given by \eqref{eq:WKB.regions}.
Notice that is is independent of $\epsilon$.
%Then, according to \eqref{eq:WKB.regions}, the relevant part of the solution is
%% 
%\bea\label{eq:WKB2tp}
%\phi_{E J}^{ \sf WKB}(r) &=& \left\{\begin{array}{lcr} 
%L^{(1)}_{EJ+}\; u_{EJ}^+(r; r_1) + L^{(1)}_{EJ-}\; 
%u_{EJ}^-(r; r_1)\,,\qquad&~&\qquad 0<r\ll r_1\,, ~~~\cr
%&&\cr
%R^{(2)}_{EJ+}\; u_{EJ}^+(r; r_2) + R^{(2)}_{EJ-}\; u_{EJ}^-(r; r_2)\,,\qquad&~&\qquad r_2\ll r<\infty\,,~~~
%\end{array}\right. 
%\eea
%%
%where  the possible $\epsilon$-dependence of the coefficients is included in the $J$-label, and 
%%
%\be\label{eq:WKB.wavespot2p}
%u_{EJ}^\pm (r; r_i) = |V_{EJ}(r)|^{-\frac{1}{4}}\;
%e^{\pm\,{\sf m}L\int_{r_i}^r dr\,\sqrt{V_{EJ}(r)}}\,,\qquad \qquad i=0,1
%\ee

%These are the functions to be used, in this approximation, to get the bi-spinors \eqref{eq:spinorphi}.
%
From the behaviors \eqref{eq:behaV} it is straightfordward to see that $u_\pm (r; r_1)|_{r\to 0} 
\propto r^{\frac{1}{2}\pm {\sf m}L\,|J|}$. 
%
%
%\bea\label{eq:upmasymptotics}
%u_{EJ}^\pm (r; r_1) 
%\propto
%&\stackrel{r\rightarrow 0}\longrightarrow& 
%B^\pm_{EJ}\;
%r^{\frac{1}{2}\pm {\sf m}L\,|J|}%\sim  r^{\pm \gamma\,\left(N+\frac{d-1\pm 1}{2}\right)}
%\quad\mbox{with}\quad
%B^\pm_{EJ}=\frac{e^{\pm {\sf m}L\,|J|\,\int_0^{r_1}dr'\ln\frac{r'}{r_1}\left(\frac{r'}{J}\sqrt{{V_{EJ}}}\right)'}}{r_1{}^{\pm{\sf m}L\,|J|}\,|J|^\frac{1}{2}}~~~
%\cr&&\cr
%u_{\omega J}^\pm (r; r_2)
%&\stackrel{r\rightarrow\infty}\longrightarrow&C^\pm_{\omega J}\,r^{\frac{1}{2} \pm\gamma}\sim r^{\pm\gamma}\crC^\pm_{\omega J} &=& \frac{e^{\pm \gamma\,\int_{r_1}^\infty \frac{dr}{r}\left(\sqrt{r^2 V_{\omega J}}-1\right)}}{r_1{}^{\pm\gamma}}
%\eea
%
Then, smoothness at the origin enforces %us to impose  
the condition $ L^{(1)}_{-}=0$ and \eqref{eq:WKB.matrix.V} becomes
\small
\be
\label{eq:L-=0}
R^{(2)}_{+} =	2\cos\left({\sf m}L\int_{r_1}^{r_2} dr'\sqrt{-V_{EJ}(r')}\right)L^{(1)}_{+},
\quad\ 
%\\
%\label{eq:L-=02}
R^{(2)}_{-} 
%&=&	
=
-\sin\left({\sf m}L\int_{r_1}^{r_2} dr'\sqrt{-V_{EJ}(r')}
\right)L^{(1)}_{+}.~~~~~
\ee
\normalsize
Furthermore, as $r$ goes to infinity we get 
\be
\phi^{\sf WKB}\approx R_{+}^{(2)}r_\Lambda^{-{\sf m}L}e^{+\,{\sf m}L\int_{r_i}^{r_\Lambda} dr\,\sqrt{V_{EJ}(r)}}\,r^{\frac12+{\sf m}L}
+
R_{-}^{(2)}r_\Lambda^{{\sf m}L}e^{-\,{\sf m}L\int_{r_i}^{r_\Lambda} dr\,\sqrt{V_{EJ}(r)}}\, r^{\frac12-{\sf m}L}\,,
\label{eq:culo2}
\ee 
Where $r_\Lambda$ is an UV cutoff. From this expression, we can read the coefficients ${A}_{\omega jm\epsilon}$ and ${B}_{\omega jm\epsilon}$ on equation \eqref{eq:culoculo}. Notice the difference in the power between the two expressions, which is not relevant in the limit of large ${\sf m}L$. 

If we search for normal modes, we have to impose normalizability conditions and then smoothness at the boundary implies that $R^{(2)}_{ +}=0$ must hold. 
From  \eqref{eq:L-=0}, this yields the Bohr-Sommerfeld quantization condition
\be 
\left.{\sf m}L\int_{r_1}^{r_2} dr'\,\sqrt{-V_{E J}(r')}\;\right|_{E=E_n(J)} 
= \left(n + \frac{1}{2}\right)\,\pi\,,
\quad \quad n\in\mathbb{N}_0\,,
\label{eq:normal.modes}
\ee 
where $E_n(J)$ is the resulting dispersion relation of the  $n$-th band.

If instead we are interested in the connected fermionic two-point correlator, with the help of \eqref{eq:culo2} we write the equations  \eqref{eq:propagators} as
%
%\bea\label{eq:propagators.WKB}
%{\cal G}_{EJ+} &=& \left(\frac{2}{E+J}\right)^2{\cal G}_{EJ-} =
%\nonumber\\
%&=&-
% \frac{L^3 r_2^{2{\sf m}L}}{E+J} \,
%e^{-2{\sf m}L\int_{r_2}^{\infty} \frac {dr}r\left(
%\sqrt{r^2V_{EJ}(r)}-1\right)}
%\,\tan\!
%\left(
%{\sf m}L\int_{r_1}^{r_2}\! dr' \sqrt{-V_{EJ}(r')}
%\right)\,.~~~~~
%\eea
%
\bea\label{eq:propagators.WKB}
{\cal G}_{EJ-} &=& \frac14({E+J})^2{\cal G}_{EJ+} =
\nonumber\\
&=&-
 \frac{L^3 }{4}r_\Lambda^{-2{\sf m}L}(E+J) \,
e^{-2{\sf m}L\int_{r_2}^{r_\Lambda}  {dr}  
\sqrt{ V_{EJ}(r)}}
\,\tan\!
\left(
{\sf m}L\int_{r_1}^{r_2}\! dr' \sqrt{-V_{EJ}(r')}
\right)\,.~~~~~
\eea
This is the formula for the correlator that we use in the main text. 

Notice that, as expected, the correlator \eqref{eq:propagators.WKB} has poles at the normal modes \eqref{eq:normal.modes}. The residua  around those poles $r_n(J)$ are obtained by multiplying the correlator by $E-E_n(J)$ and then taking the limit $E\approx E_n(J)$. We get the expression
\begin{equation}
r_n(J)= \frac{L^2 }{4{\sf m}}\,\frac{E_n(J)+J}{E_n(J)} \,
\frac{e^{-2{\sf m}L\int_{r_2}^{r_\Lambda}  {dr}  
\sqrt{ V_{EJ}(r)}}}{
\int_{r_1}^{r_2} dr'\,\frac{ e^{\lambda(r)-\nu(r)}}{\sqrt{-V_{E_n(J)J}}}}\,r_\Lambda^{-2{\sf m}L}\,.
\label{eq:residua}
\end{equation}
This implies that, close to any pole, a good approximation of the correlator is given by the expression
\begin{equation}
{\cal G}_{EJ-}\sim \frac{r_n(J)}{E-E_n(J)}\,.
\end{equation}
With this, we can define a \emph{regular part} of the correlator according to
\begin{equation}
{\cal G}_{EJ-}^{\sf reg}={\cal G}_{EJ-}-\sum_n\frac{r_n(J)}{E-E_n(J)}\,.
\end{equation}

\newpage

\eject

 \end{document}